\documentclass[12pt]{article}
\usepackage{amsfonts,amsmath,amssymb,amsthm}
\newtheorem{theo}{Theorem}
\newtheorem{prop}{Proposition}
\newtheorem{lemm}{Lemma}
\newtheorem{coro}{Corollary}
\def\real{{\mathbb R}}
\def\integer{{\mathbb Z}}
\theoremstyle{remark}
\newtheorem{rem}{Remark}
\newtheorem{example}{Example}
\theoremstyle{definition}
\newtheorem{defi}{Definition}

\date{}
\author{I.\,A.\,Dynnikov\footnote{Mech.\ and Math.\ Dept.,
Moscow State University, Moscow 119992 GSP-2, Russia,
e-mail:~dynnikov@mech.math.msu.su},
S.\,P.\,Novikov\footnote{Institute for Physical Sciences and
Technology, Univ.\ of Maryland at College Park, MD 20742, USA;
L.\,D.\,Landau Institute for Theoretical Physics, Kosygyna str.~2,
Moscow 117940, Russia, e-mail:~novikov@ipst.umd.edu}}
\title{Geometry of the Triangle Equation on Two-Manifolds\footnote{The work of
I.\,Dynnikov is partially supported by Russian Foundation for
Fundamental Research, grant no.~02-01-00659, and by
Leading Scientific School Support grant no.~00-15-96011; the work of
S.\,Novikov is supported by NSF (USA), grant no.~DMS-00772700.
Author's contributions are equal; author's names are listed
in the alphabetical order.}}

\begin{document}
\maketitle
\vbox to0pt{\vss\hbox to\textwidth{\hss\it Dedicated to V.\,I.\,Arnol'd's 65th
birthdate}%
\vskip6cm}

{\abstract{A non-traditional approach to the discretization
of differential-geometrical
connections was suggested by the authors in 1997. At the same
time we started studying first order difference ``black and white
triangle operators (equations)'' on triangulated surfaces with
a black and white coloring of triangles.
In this work, we develop the theory
of these operators and equations, showing their similarity with
the complex derivatives $\partial$ and $\overline\partial$.}}

\section*{Introduction}

Do there exist any natural difference analogs of the operators
$\partial$ and $\overline\partial$ on the complex plane? No theory
of difference ``first order'' operators that have some properties
similar to those of $\partial$ and $\overline\partial$ has been developed
before; the attention of the literature is paid only to discrete
analogs of the Laplace--Beltrami operator, which are ``second
order'' operators. However, in papers \cite{nd,dn,n1} we
already studied some ``first order'' difference operators,
starting from the problem of finding a discrete analog of the
Laplace transformation for a standard two-dimensional second
order linear differential equation. Our operators were originally
defined on the equilateral triangular lattice in the space
$\real^2$ (see \cite{n1}). In \cite{nd,dn} these definitions were
extended to arbitrary ``black and white'' triangulations
of two-dimensional surfaces without boundary.  The ``black (white)
triangle'' equations, which have naturally appeared in the theory,
are studied here with respect to their analogy with the operators
$\partial$ and $\overline\partial$. It turns out that their
properties partially imitate some properties of complex
derivatives on two-dimensional manifolds.

\section{First order triangle operators on triangulated
surfaces. Discrete connections}\label{disconn}

Let $M$ be a two-dimensional manifolds without boundary
endowed with a triangulation $\cal T$. Let $\cal K$
be a family of triangles from $\cal T$ such that,
for any vertex $P$ of $\cal T$, there is
at least one triangle $T\in\cal K$ adjacent
to $P$. Let us fix some numeric coefficients
$b_{T,P}$ defined for any triangle $T\in\cal K$
and its vertex $P\in T$.

\begin{defi}
By a \emph{first order (or triangle) operator} on $M$ associated
with the family $\cal K$ and coefficients $b_{T,P}$ we call an
operator $Q^{\cal K}$ defined by the formula
\begin{equation}\label{1order}
(Q^{\cal K}\psi)_T=\sum\limits_{P\in T}b_{T,P}\psi_P.
\end{equation}
The operator $Q^{\cal K}$ is well-defined as a linear map of the
space of all functions $\psi_P$ of a vertex $P$ into the space
of functions $\varphi_T=(Q^{\cal K}\psi)_T$ of a triangle $T$ from
the selected family  $\cal K$.
\end{defi}

\begin{rem}
In 1997 (see \cite{nd}) we considered  operators on
multi-dimensional manifolds as well, but in this paper we focus
only on the two-dimensional case, where an analogy with the
operators $\partial$, $\overline\partial$ takes place.

In Appendix~2 we provide a generalization of the constructions
of this section to higher dimensions.
\end{rem}

\noindent{\bf The maximal family: discrete
connections}. In the case of the maximal
family $\cal K$ equal to the set of all triangles of the
triangulation $\cal T$, the theory of the equation
\begin{equation}\label{geneq}
Q^{\cal K}\psi=0
\end{equation}
naturally leads to the notions of a \emph{discrete connection} and
\emph{curvature}. The coefficients $b_{T,P}$ play the role of
the connection coefficients. Following \cite{nd}, we now introduce
the \emph{local holonomy} (or the \emph{curvature}) of the connection
$\{b_{T,P}\}$ at a vertex $P$, which is an operator
$\real^2\rightarrow\real^2$ defined by a matrix of the form
\begin{equation}
K_P=\begin{pmatrix} 1&k'_P\\0&k_P''\end{pmatrix}.
\end{equation}
It is constructed as follows.

Let us enumerate all the vertices and triangles included
in the star of a vertex $P$ as shown in Fig.~\ref{obxod}.
\begin{figure}[ht]\caption{}\label{obxod}
\vskip10pt \centerline{\begin{picture}(100,100)
\put(50,50){\circle{3}}\put(90,50){\circle{3}}\put(70,85){\circle{3}}
\put(30,85){\circle{3}}\put(10,50){\circle{3}}\put(30,15){\circle{3}}
\put(70,15){\circle{3}}\put(52,50){\line(1,0){36}}\put(12,50){\line(1,0){36}}
\bezier200(50.8,51.4)(60,67.5)(69.2,83.6)
\bezier200(10.8,51.4)(20,67.5)(29.2,83.6)
\bezier200(70.8,16.4)(80,32.5)(89.2,48.6)
\bezier200(30.8,16.4)(40,32.5)(49.2,48.6)
\bezier200(49.2,51.4)(40,67.5)(30.8,83.6)
\bezier200(89.2,51.4)(80,67.5)(70.8,83.6)
\bezier200(29.2,16.4)(20,32.5)(10.8,48.6)
\bezier200(69.2,16.4)(60,32.5)(50.8,48.6)
\put(32,85){\line(1,0){36}} \put(50,15){\hbox
to0pt{\hss$\dots$\hss}} \put(70,61){\vbox to0pt{\vss\hbox
to0pt{\hss$T_1$\hss}\vss}} \put(50,73){\vbox to0pt{\vss\hbox
to0pt{\hss$T_2$\hss}\vss}} \put(30,61){\vbox to0pt{\vss\hbox
to0pt{\hss$T_3$\hss}\vss}} \put(30,38){\vbox to0pt{\vss\hbox
to0pt{\hss$T_4$\hss}\vss}} \put(70,38){\vbox to0pt{\vss\hbox
to0pt{\hss$T_n$\hss}\vss}} \put(94,50){\vbox to0pt{\vss\hbox
to0pt{$P_1$\hss}\vss}} \put(6,50){\vbox to0pt{\vss\hbox
to0pt{\hss$P_4$}\vss}} \put(72,90){\vbox to0pt{\vss\hbox
to0pt{\hss$P_2$\hss}}} \put(28,90){\vbox to0pt{\vss\hbox
to0pt{\hss$P_3$\hss}}} \put(72,11){\vbox to0pt{\hbox
to0pt{\hss$P_n$\hss}\vss}} \put(28,11){\vbox to0pt{\hbox
to0pt{\hss$P_5$\hss}\vss}} \put(50,42){\vbox to0pt{\hbox
to0pt{\hss$P$\hss}\vss}}
\end{picture}}
\end{figure}
We will be constructing a local solution of
equation~(\ref{geneq}) in the star of $P$,
starting from arbitrarily chosen values
of $\psi$ at vertices $P,P_1$. We will have
$$\begin{aligned}
\psi_{P_2}&=-\frac1{b_{T_1,P_2}}(b_{T_1,P}\psi_P+b_{T_1,P_1}\psi_{P_1}),\\
\psi_{P_3}&=-\frac1{b_{T_2,P_3}}(b_{T_2,P}\psi_P+b_{T_2,P_2}\psi_{P_2}),\\
&\dots\\
\widetilde
\psi_{P_1}&=-\frac1{b_{T_n,P_1}}(b_{T_n,P}\psi_P+b_{T_n,P_n}\psi_{P_n})=
k_P'\psi_P+k_P''\psi_{P_1},
\end{aligned}$$
where
$$\begin{aligned}
k_P'&=\sum\limits_{k=0}^{n-1}(-1)^{k+1}\;
\frac{b_{T_n,P}\cdot\ldots\cdot b_{T_{n-k},P}}
{b_{T_n,P_1}\cdot\ldots\cdot b_{T_{n-k},P_{n-k+1}}}\\
k_P''&=(-1)^n\;\frac{b_{T_1,P_1}b_{T_2,P_2}\cdot\ldots\cdot b_{T_n,P_n}}
{b_{T_1,P_2}b_{T_2,P_3}\cdot\ldots\cdot b_{T_n,P_1}}.
\end{aligned}$$
(In the formula for $k_P'$, we assume $P_{n+1}=P_1$.)

Thus, having started from arbitrary values $\psi_P,\psi_{P_1}$,
after a full turn around the vertex $P$, we come to new ones
$$(\widetilde\psi_P,\widetilde\psi_{P_1})=
(\psi_P,\psi_{P_1})\begin{pmatrix} 1&k'_P\\0&k_P''\end{pmatrix}.$$
We say that the connection $\{b_{T,P}\}$ has  \emph{zero
curvature} at the vertex $P$ if, for any $\psi_P,\psi_P'$, we have
$(\widetilde\psi_P,\widetilde\psi_{P_1})= (\psi_P,\psi_{P_1})$,
{\it i.e.}, if we have $$k_P'=0,\quad k_P''=1.$$

\begin{example}
Assume that all the connection coefficients are equal to one,
$b_{T,P}\equiv1$, and the number $n_P$ of triangles
adjacent to a vertex $P$ is even, $n_P=2l_P$.
Then the holonomy of the connection $\{b_{T,P}\}$ at the
vertex $P$ is trivial. Indeed, we have
$$(\psi_P,\psi_{P_{k+1}})=(\psi_P,\psi_{P_k})
\begin{pmatrix}1&-1\\0&-1\end{pmatrix},\qquad
K_P=\begin{pmatrix}1&-1\\0&-1\end{pmatrix}^{n_P}=
\begin{pmatrix}1&0\\0&1\end{pmatrix}.$$
\end{example}

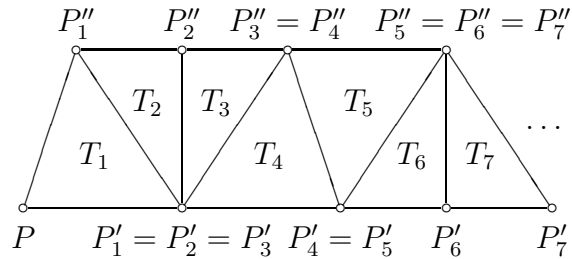
\begin{figure}[ht]\caption{A thick path}\label{thick}\vskip10pt
\centerline{\begin{picture}(220,100)
\put(10,20){\circle{3}}
\put(10,5){\hbox to0pt{\hss$P$\hss}}
\put(70,20){\circle{3}}
\put(70,5){\hbox to0pt{\hss$P_1'=P_2'=P_3'$\hss}}
\put(30,80){\circle{3}}
\put(30,87){\hbox to0pt{\hss$P_1''$\hss}}
\put(37,40){\vbox to0pt{\vss\hbox to0pt{\hss$T_1$\hss}\vss}}
\put(12,20){\line(1,0){56}}
\put(10.5,21.5){\line(1,3){19}}
\put(69,21.5){\line(-2,3){38}}
\put(70,80){\circle{3}}
\put(70,87){\hbox to0pt{\hss$P_2''$\hss}}
\put(57,60){\vbox to0pt{\vss\hbox to0pt{\hss$T_2$\hss}\vss}}
\put(32,80){\line(1,0){36}}
\put(70,22){\line(0,1){56}}
\put(110,80){\circle{3}}
\put(110,87){\hbox to0pt{\hss$P_3''=P_4''$\hss}}
\put(83,60){\vbox to0pt{\vss\hbox to0pt{\hss$T_3$\hss}\vss}}
\put(71,21.5){\line(2,3){38}}
\put(72,80){\line(1,0){36}}
\put(130,20){\circle{3}}
\put(130,5){\hbox to0pt{\hss$P_4'=P_5'$\hss}}
\put(103,40){\vbox to0pt{\vss\hbox to0pt{\hss$T_4$\hss}\vss}}
\put(72,20){\line(1,0){56}}
\put(110.5,78.5){\line(1,-3){19}}
\put(170,80){\circle{3}}
\put(170,87){\hbox to0pt{\hss\hskip20pt$P_5''=P_6''=P_7''$\hss}}
\put(137,60){\vbox to0pt{\vss\hbox to0pt{\hss$T_5$\hss}\vss}}
\put(131,21.5){\line(2,3){38}}
\put(112,80){\line(1,0){56}}
\put(170,20){\circle{3}}
\put(170,5){\hbox to0pt{\hss$P_6'$\hss}}
\put(157,40){\vbox to0pt{\vss\hbox to0pt{\hss$T_6$\hss}\vss}}
\put(132,20){\line(1,0){36}}
\put(170,22){\line(0,1){56}}
\put(210,20){\circle{3}}
\put(210,5){\hbox to0pt{\hss$P_7'$\hss}}
\put(183,40){\vbox to0pt{\vss\hbox to0pt{\hss$T_7$\hss}\vss}}
\put(172,20){\line(1,0){36}}
\put(171,78.5){\line(2,-3){38}}
\put(200,50){$\dots$}
\end{picture}}
\end{figure}

The \emph{global holonomy} of a discrete connection $\{b_{T,P}\}$
is defined for any ``thick'' path $\gamma$, which is a sequence of
triangles $T_1,\dots,T_m$ such that, for all
$j=1,\dots,m-1$, the triangles $T_j$ and
$T_{j+1}$ have a common edge $\kappa_j=T_j\cap T_{j+1}$,
whose vertices are denoted by $P_j',P_j''$ (see
Fig.~\ref{thick}).
This is done as follows. Start from a solution
$(\psi_P,\psi_{P_1'},\psi_{P_1''})$ of equation~(\ref{geneq}) in
the triangle $T_1$:
$$b_{T_1,P}\psi_{P}+b_{T_1,P_1'}\psi_{P_1'}+b_{T_1,P_1''}=0.$$
We assume that $\psi_P$ and $\psi_{P_1'}$ are chosen arbitrarily,
and $\psi_{P_1''}$ is found from the equation.
Then we extend the solution along $\gamma$ successively
to the triangles
$T_2,T_3,\dots$, {\it i.e.}, to the
vertices $P_2',P_2'',P_3',P_3'',\dots$. This can be done in a unique way.
If $\gamma$ is a loop, {\it i.e.},~$T_{m+1}=T_1$, then,
having passed over $\gamma$, we will obtain
a new solution $(\widetilde\psi_P,\widetilde\psi_{P_1'},
\widetilde\psi_{P_1''})$ of~(\ref{geneq}) in the triangle
$T_1$. Thus, we have constructed the \emph{holonomy operator}
$R_\gamma:\real^2\rightarrow\real^2$:
$$(\psi_P,\psi_{P_1'})\mapsto(\widetilde\psi_P,\widetilde\psi_{P_1'})=
(\psi_P,\psi_{P_1'})\cdot R_\gamma.$$

\begin{lemm}\label{l1}
The global holonomy of any discrete connection $\{b_{P,T}\}$ with
zero curvature is a well defined homomorphism
$\pi_1(M,T_1)\rightarrow GL(2,\real)$.
\end{lemm}

\begin{proof}
For any two adjacent triangles $T,T'$, {\it i.e.},~ones having
exactly one common edge, we denote by $R_{T,T'}$ the extension
operator of a solution of~(\ref{geneq}) from the triangle $T$ to
$T'$: $\psi^{(T)}\mapsto\psi^{(T')}= \psi^{(T)}\cdot R_{T,T'}$. By
definition, we have $$R_\gamma=R_{T_1,T_2}\cdot\ldots\cdot
R_{T_{m-1},T_m} \cdot R_{T_m,T_{m+1}}.$$

Clearly, the correspondence $\gamma\mapsto R_\gamma$
is multiplicative:
$$R_{\gamma_1\circ\gamma_2}=R_{\gamma_1}\circ R_{\gamma_2},$$
and consistent with the inversion:
$$R_{\gamma^{-1}}=(R_\gamma)^{-1}.$$
We have only to check the invariance of $R_\gamma$
under a homotopy of the path $\gamma$.

Any homotopy of $\gamma$ can be reduced to finitely many
insertions and removals of fragments like this:
$$\begin{aligned}
\ldots,T,T',T,\ldots\quad&\leftrightarrow\quad\ldots,T,\ldots,\\
\ldots,T,T',T'',\ldots,T^{(n)},T,\ldots\quad&\leftrightarrow\quad
\ldots,T,\ldots,
\end{aligned}$$
where $T,T',\dots,T^{(n)}$ are triangles from the star
of a vertex listed in the order we meet them when
walking around the vertex. The holonomy operator $R_\gamma$
is unchanged under the first operation in view of
$R_{T,T'}R_{T',T}=\mathrm{id}$, and it does so under
the latter in view of the triviality of the local holonomy.\end{proof}

\begin{theo}\label{anyrho}
For any triangulation $\cal T$ of a surface $M$ and any
homomorphism $\rho:\pi_1(M)\rightarrow GL(2,\real)$ there
exists a zero curvature discrete connection $\{b_{T,P}\}$
that generates the representation $\rho$.
\end{theo}

See Appendix~1 for a proof of this theorem.

\medskip
In the sequel, we will consider in detail only
one special case in which the star of each vertex of $\cal T$
consists of an even number of triangles and
all the connection coefficients are equal to $1$,
$b_{T,P}\equiv1$. As we have already seen,
in this case, the connection $\{b_{T,P}\}$ has zero curvature,
and hence, defines a homomorphism
$\rho:\pi_1(M)\rightarrow GL(2,\real)$.
We call this connection \emph{canonical}.

\begin{prop}\label{p1}
The holonomy group of the canonical connection
associated with a triangulation all whose vertices
have an even valence is a subgroup of
the group $S_3\subset GL(2,\real)$ generated by
matrices
\begin{equation}
\begin{pmatrix}0&1\\1&0\end{pmatrix},\qquad
\begin{pmatrix}-1&0\\-1&1\end{pmatrix}.
\end{equation}
\end{prop}

\begin{proof}
Let $\gamma$ be a thick path $T_1,T_2,\dots,T_{m+1}=T_1$. Let us color
the vertices of $T_1$ into three distinct colors, say,
$a,b,c$, and then continue the coloring so that all three vertices
of any triangle $T_j$, $j=2,3,\dots$, have
different colors (see Fig.~\ref{color}).
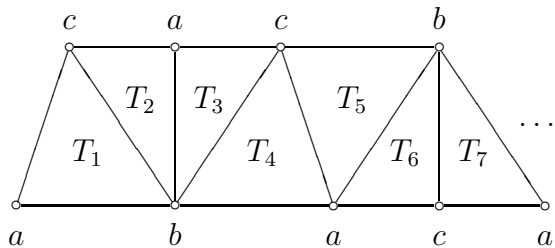
\begin{figure}[ht]\caption{Coloring vertices in three colors}
\label{color}\vskip10pt
\centerline{\begin{picture}(220,100)
\put(10,20){\circle{3}}
\put(10,5){\hbox to0pt{\hss$a$\hss}}
\put(70,20){\circle{3}}
\put(70,5){\hbox to0pt{\hss$b$\hss}}
\put(30,80){\circle{3}}
\put(30,87){\hbox to0pt{\hss$c$\hss}}
\put(37,40){\vbox to0pt{\vss\hbox to0pt{\hss$T_1$\hss}\vss}}
\put(12,20){\line(1,0){56}}
\put(10.5,21.5){\line(1,3){19}}
\put(69,21.5){\line(-2,3){38}}
\put(70,80){\circle{3}}
\put(70,87){\hbox to0pt{\hss$a$\hss}}
\put(57,60){\vbox to0pt{\vss\hbox to0pt{\hss$T_2$\hss}\vss}}
\put(32,80){\line(1,0){36}}
\put(70,22){\line(0,1){56}}
\put(110,80){\circle{3}}
\put(110,87){\hbox to0pt{\hss$c$\hss}}
\put(83,60){\vbox to0pt{\vss\hbox to0pt{\hss$T_3$\hss}\vss}}
\put(71,21.5){\line(2,3){38}}
\put(72,80){\line(1,0){36}}
\put(130,20){\circle{3}}
\put(130,5){\hbox to0pt{\hss$a$\hss}}
\put(103,40){\vbox to0pt{\vss\hbox to0pt{\hss$T_4$\hss}\vss}}
\put(72,20){\line(1,0){56}}
\put(110.5,78.5){\line(1,-3){19}}
\put(170,80){\circle{3}}
\put(170,87){\hbox to0pt{\hss$b$\hss}}
\put(137,60){\vbox to0pt{\vss\hbox to0pt{\hss$T_5$\hss}\vss}}
\put(131,21.5){\line(2,3){38}}
\put(112,80){\line(1,0){56}}
\put(170,20){\circle{3}}
\put(170,5){\hbox to0pt{\hss$c$\hss}}
\put(157,40){\vbox to0pt{\vss\hbox to0pt{\hss$T_6$\hss}\vss}}
\put(132,20){\line(1,0){36}}
\put(170,22){\line(0,1){56}}
\put(210,20){\circle{3}}
\put(210,5){\hbox to0pt{\hss$a$\hss}}
\put(183,40){\vbox to0pt{\vss\hbox to0pt{\hss$T_7$\hss}\vss}}
\put(172,20){\line(1,0){36}}
\put(171,78.5){\line(2,-3){38}}
\put(200,50){$\dots$}
\end{picture}}
\end{figure}
At the same time, we extend a solution $\psi$ from the
triangle $T_1$ to $T_2,T_3,\dots$.
Since the sum of the values
of $\psi$ over the vertices of any triangle must be zero,
we will have $\psi_{P}=\psi_{P'}$ for any two vertices
$P,P'$ of the same color. We denote by $\psi_a,\psi_b,\psi_c$
the values of $\psi$ at vertices of the corresponding color.
We have $\psi_a+\psi_b+\psi_c=0$.

After passing over $\gamma$, we come to another coloring of the
vertices of $T_1$, which is obtained from the original one by a
permutation from $S_3$. The group $S_3$ is generated by the
permutations $\begin{pmatrix} a&b&c\\b&a&c\end{pmatrix}$ and
$\begin{pmatrix} a&b&c\\c&b&a\end{pmatrix}$. For the first
permutation, we have
$$(\widetilde\psi_a,\widetilde\psi_b)=(\psi_b,\psi_a)=(\psi_a,\psi_b)\cdot
\begin{pmatrix}0&1\\1&0\end{pmatrix},$$ for the latter one, we have
$$(\widetilde\psi_a,\widetilde\psi_b)=(-\psi_a-\psi_b,\psi_b)=
(\psi_a,\psi_b)\cdot\begin{pmatrix}-1&0\\-1&1\end{pmatrix}.$$
\hfill\end{proof}

In the case of $\{b_{T,P}\}$ the canonical connection,
we call equation~(\ref{geneq}) the \emph{triangle equation}:
\begin{equation}\label{triangleeq}
(Q\psi)_T=\sum\limits_{P\in T}\psi_P=0.
\end{equation}
In order to construct a solution, we have to take initial values
$\psi_a,\psi_b$ invariant under all holonomy operators. According
to Proposition~\ref{p1}, there are three cases:
\begin{description}
\item[Case 1]: the holonomy group $G$ is isomorphic either
to $S_3$ or $S_3^+$ (the group of even permutations);
the representation $G\rightarrow GL(2,\real)$ is irreducible;
\item[Case 2]: the holonomy group is isomorphic to $\integer_2$;
up to multiplicative constant, there is exactly one
invariant vector;
\item[Case 3]: the holonomy group is trivial;
all vectors from $\real^2$ are invariant.
\end{description}
Thus, we have proved the following.

\begin{theo}
The space of ``covariant constants'', i.e., of solutions of the
triangle equation, is:
\begin{description}
\item two-dimensional if the holonomy group is trivial;
\item one-dimensional if the holonomy group is isomorphic to $\integer_2$;
\item trivial in the other cases.
\end{description}
\end{theo}

Now consider the operator $L=Q^+Q$, where $Q$ is the first
order operator associated with the canonical connection
on the surface $M$. The operator $L$ can be written
as follows:
\begin{equation}
L=-2\delta d+3n_P=-2\Delta+3n_P,
\end{equation}
where $n_P$ is the ``potential'' equal
to the valence of a vertex,
$d,\delta$ are the boundary and coboundary operators,
respectively, $d=\delta^+$.

\begin{prop}\label{p2}
The operator $L$ has a non-trivial space of zero modes $L\psi=0$
if and only if the holonomy
group of the canonical connection is either trivial, in which case
the space of zero modes is two-dimensional, or isomorphic to
$\integer_2$, in which case it is one-dimensional. These modes
$\psi$ satisfy the equation $Q\psi=0$.
\end{prop}

\begin{rem}
For $\cal T$ a \emph{uniform} triangulation, {\it i.e.},~such that
all vertices have the same valence $n_P=2l=\mathrm{const}$, the
zero modes of $L$ are eigenfunctions of the Laplace--Beltrami
operator $\Delta$ with maximal eigenvalue.
\end{rem}

It is interesting to compare our constructions with some results
previously obtained in graph theory (see \cite{s}). Let us
consider the Poincare dual graph  $\Gamma$ of our triangulation.
Vertices of $\Gamma$ correspond to triangles of $\cal T$.
This graph $\Gamma$
is ``dichromatic'' in the graph-theoretical terminology if and
only if there exists a coloring of triangles  of $\cal T$
in black and white such that any two adjacent triangles
have different colors. The Laplace--Beltrami operators on dichromatic
graphs $\Gamma$ were studied in~\cite{s}. The space ${\cal L}_\Gamma$ of
functions depending on vertices of $\Gamma$
splits into the direct sum
$${\cal L}_{\Gamma}={\cal L}_{\mathrm b}\oplus{\cal L}_{\mathrm w},$$
where  ${\cal L}_{\mathrm b}$ (respectively, ${\cal L}_{\mathrm w}$) is
the space of functions of black (respectively, white) triangles of $\cal T$.

In \cite{nd} we studied operators $Q_{\mathrm{bw}}:{\cal L}_{\mathrm w}
\rightarrow{\cal L}_{\mathrm b}$,
and the operators $Q_{\mathrm{wb}}=Q^+_{\mathrm{bw}}$ adjoint to them,
associated with the coloring. They were
needed for a generalization of the Laplace
transformation. This was a new type of first order difference
operators, except in the following case: for an equilateral
triangular lattice with the natural coloring, both spaces
${\cal L}_{\mathrm w},{\cal L}_{\mathrm b}$ are canonically
isomorphic to the space of functions
depending on vertices (see below). In our previous work \cite{nd},
``second order Schr\"odinger operators'' of the form
$L=Q_{\mathrm{wb}}Q^+_{\mathrm{wb}}$
and $L'=Q^+_{\mathrm{wb}}Q_{\mathrm{wb}}$ acting on the spaces
${\cal L}_{\mathrm w}$ and ${\cal L}_{\mathrm b}$,
respectively, were discussed.

It is easy to see that the standard graph-theoretical
Laplace--Beltrami operator on the dual graph $\Gamma$ has a matrix
of the form:
$$\Delta_{\Gamma}=\begin{pmatrix}0&Q_{\mathrm{wb}}\\
Q_{\mathrm{wb}}^+&0\end{pmatrix},\qquad
\Delta^2_{\Gamma}=\begin{pmatrix}Q_{\mathrm{wb}}Q_{\mathrm{wb}}^+
&0\\0&Q_{\mathrm{wb}}^+Q_{\mathrm{wb}}
\end{pmatrix}=L\oplus L'.$$

So the results obtained in the work of P.\,Sarnak~\cite{s} involve,
in fact, the operators of the type $Q_{\mathrm{wb}}$ but not of the type
studied in the present work.

Let us ask the following

\smallskip
\noindent{\bf Question:} when a triangulation $\cal T$ of a surface $M$
admits a black and white (or b/w) coloring of triangles such that
any two adjacent triangles are of different colors?

The answer, which is obvious, is given in the following Lemma.

\begin{lemm}\label{l2}
A b/w coloring of a triangulation $\cal T$
exists if and only if any thick loop
$\gamma=(T_1,\dots,T_m,T_{m+1}=T_1)$, where $T_i\cap T_{i+1}$ is an
edge for all $i=1,\dots,m$, has an even length: $m=2k$.
\end{lemm}

Thus, we have the following three homomorphisms
$\rho_1,\rho_2,\rho_3:\pi_1(M)\rightarrow\integer_2$:
\def\labelenumi{\theenumi)}
\begin{enumerate}
\item the parity $\rho_1$ of the global holonomy, $\pi_1(M)
\rightarrow S_3\rightarrow\integer_2$;
\item the orientation homomorphism $\rho_2$;
\item the parity homomorphism $\rho_3$ of the number of
triangles in a thick loop.
\end{enumerate}

\begin{lemm}\label{l3}
We have $\rho_1\rho_2=\rho_3$.
\end{lemm}

We skip the easy proof.

\begin{coro}
a) Let $M$ be orientable; then a b/w coloring of $\cal T$ exists
if and only if the holonomy group of the canonical connection
associated with $\cal T$ is either trivial or isomorphic to the
group $S_3^+$ of even permutations;

b) Let $M$ be non-orientable; then a b/w coloring exists if and
only if the holonomy group is isomorphic to $\integer_2$ and we
have $\rho_1=\rho_2$.
\end{coro}

In the case $\rho_3=1$, we fix a b/w coloring of $\cal T$ and
consider the first order operators $Q_{\mathrm b}$ and $Q_{\mathrm
w}$ of the form~(\ref{1order}) associated with the families of
black and white triangles, respectively. The corresponding
equations $Q_{\mathrm b}\psi=0$ and $Q_{\mathrm w}\psi=0$ are
called \emph{black triangle equation} and \emph{white triangle
equation}, respectively. Clearly, we have
\begin{equation}
Q_{\mathrm b}^+Q_{\mathrm b}=-\delta d+\frac32\,n_P=
Q_{\mathrm w}^+Q_{\mathrm w}.
\end{equation}

\section{Maximum principle}\label{mp}
As before, let $M$ be a surface endowed with a triangulation $\cal
T$. Let $M'\subset M$ be a surface (with boundary) that has a form
of a finite simplicial subcomplex of $M$. Assume that $M'$ is
orientable and the holonomy of the canonical connection of $M'$ is
globally trivial.

As we saw in the previous Section,  there exists a b/w coloring of
triangles from $M'$ such that any two adjacent triangles have
different colors. Also, there exists a coloring of vertices into
three colors, $a,b,c$, such that all three vertices of any
triangle are of different colors. The space $V$ of covariant
constants is two-dimensional and consists of functions $\psi_P$
that depend only on the color of a vertex $P$, and the values
$\psi_a,\psi_b,\psi_c$ of $\psi$ at vertices of the corresponding
color satisfy the relation $\psi_a+\psi_b+\psi_c=0$.

Let $\cal K$ be the set of black triangles in $M'$.
For any solution $\psi$ of the black triangle
equation $Q^{\cal K}\psi=0$, we define a mapping
$\widehat\psi:{\cal K}\rightarrow V\cong\real^2$ from the set
of black triangles to the space of covariant constants
on $M'$ as follows. For a black triangle $T\subset M'$, we
define $\widehat\psi(T)$ to be the covariant constant that
coincides with $\psi$ on vertices of $T$. In other words,
$\widehat\psi(T)=(\psi_{P_a(T)},\psi_{P_b(T)},\psi_{P_c(T)})$,
where $P_a(T),P_b(T),P_c(T)$ are vertices of $T$
of colors $a,b,c$, respectively.

Let us denote by $\partial_-M'$ the set of triangles in $M'$
that are adjacent to the boundary $\partial M'$, {\it i.e.},~such
that $T\subset M'$ and $T\cap\partial M'\ne\varnothing$.

\begin{theo}[Maximum principle]\label{max}
For any solution $\psi$ of the black triangle equation on a
submanifold $M'\subset M$, the image $\widehat\psi({\cal K})$ of
the set of black triangles in the space of covariant constants is
contained in the convex hull of the image of the boundary
$\widehat\psi(\partial_-M')$.
\end{theo}

\begin{proof}
We assume that $\psi$ is not a covariant constant, in which
case the assertion is trivial because the image
$\widehat\psi({\cal K})$ is just one point $\psi\in V$.

Since the complex $M'$ is finite, the convex hull $C$ of
$\widehat\psi({\cal K})$ is a polygon in the plane
$V\cong\real^2$. The assertion means that the corners of $C$ are
the images of boundary triangles $T\in{\cal K}\cap(\partial_-M')$.
In order to prove this, we are going to show that the image of any
internal black triangle, {\it i.e.},~a triangle $T\in{\cal
K}\backslash(\partial_-M')$, is not a corner of $C$.

For an internal black triangle $T\subset M'$, let us denote
by $T_1,\dots,T_6$ the six black triangles that are at distance~$2$
from $T$ in the sense of thick paths and are
arranged as shown in Fig.~\ref{star}. (The pairs of triangles
$(T_1,T_6)$, $(T_2,T_3)$, $(T_4,T_5)$ may coincide, which
will not affect the proof.)
\begin{figure}[ht]\caption{}\label{star}\vskip10pt
\centerline{\begin{picture}(150,135)
\put(45,0){\circle{3}}\put(47,1){\line(2,1){26}}\put(45,2){\line(0,1){41}}
\put(105,0){\circle{3}}\put(103,1){\line(-2,1){26}}\put(105,2){\line(0,1){41}}
\put(75,15){\circle{3}}\put(76,16){\line(1,1){28}}\put(74,16){\line(-1,1){28}}
\put(45,45){\circle{3}}\put(47,45){\line(1,0){56}}\put(46,47){\line(1,2){28}}
\put(44.5,46.5){\line(-1,3){14}}\put(43.5,45.5){\line(-3,1){42}}
\put(105,45){\circle{3}}\put(104,47){\line(-1,2){28}}
\put(105.5,46.5){\line(1,3){14}}\put(106.5,45.5){\line(3,1){42}}
\put(0,60){\circle{3}}\put(1,61){\line(1,1){28}}\put(150,60){\circle{3}}
\put(149,61){\line(-1,1){28}}\put(30,90){\circle{3}}
\put(31.5,90.5){\line(3,1){42}}\put(30,92){\line(0,1){41}}
\put(120,90){\circle{3}}\put(118.5,90.5){\line(-3,1){42}}
\put(120,92){\line(0,1){41}}\put(75,105){\circle{3}}
\put(73.5,106){\line(-3,2){42}}\put(76.5,106){\line(3,2){42}}
\put(30,135){\circle{3}}\put(120,135){\circle{3}}
\put(75,65){\vbox to0pt{\vss\hbox to0pt{\hss$T$\hss}\vss}}
\put(55,20){\vbox to0pt{\vss\hbox to0pt{\hss$T_5$\hss}\vss}}
\put(95,20){\vbox to0pt{\vss\hbox to0pt{\hss$T_6$\hss}\vss}}
\put(125,65){\vbox to0pt{\vss\hbox to0pt{\hss$T_1$\hss}\vss}}
\put(105,110){\vbox to0pt{\vss\hbox to0pt{\hss$T_2$\hss}\vss}}
\put(45,110){\vbox to0pt{\vss\hbox to0pt{\hss$T_3$\hss}\vss}}
\put(25,65){\vbox to0pt{\vss\hbox to0pt{\hss$T_4$\hss}\vss}}
\put(110,42){\vbox to0pt{\hbox to0pt{$a$\hss}\vss}}
\put(40,42){\vbox to0pt{\hbox to0pt{\hss$c$}\vss}}
\put(75,112){\vbox to0pt{\vss\hbox to0pt{\hss$b$\hss}}}
\put(75,8){\vbox to0pt{\hbox to0pt{\hss$b$\hss}\vss}}
\put(125,92){\vbox to0pt{\vss\hbox to0pt{$c$\hss}}}
\put(25,92){\vbox to0pt{\vss\hbox to0pt{\hss$a$}}}
\end{picture}}
\end{figure}
If $T$ is an internal triangle of $M'$, then all six
triangles $T_1,\dots,T_6$ lie in $M'$. Without loss
of generality we may assume that vertices are
colored as showing in Fig.~\ref{star}.

There are three families of parallel straight lines in the space
of covariant constants: $\psi_a=\mathrm{const}$,
$\psi_b=\mathrm{const}$, and $\psi_c=\mathrm{const}$.
If two black triangles $T,T'\subset M'$ have
a common vertex of color $x\in\{a,b,c\}$, then
the points $\widehat\psi(T),\widehat\psi(T')\in V$
lie in a straight line $\psi_x=\mathrm{const}$.
In Fig.~\ref{convexhull}, one of possible arrangements
of points $\widehat\psi(T),\widehat\psi(T_1),\dots,
\widehat\psi(T_6)$, which are marked simply as $0,1,2,3,4,5,6$,
respectively, is shown.
\begin{figure}[ht]\caption{}\label{convexhull}\vskip10pt
\centerline{\begin{picture}(195,135)
\put(10,20){\circle*{3}}
\put(10,26){\vbox to0pt{\vss\hbox to0pt{\hss$1$}}}
\put(0,20){\line(1,0){150}}
\put(152,20){\vbox to0pt{\vss\hbox to0pt{$\psi_a=\mathrm{const}_1$\hss}}}
\put(90,20){\circle*{3}}
\put(90,15){\vbox to0pt{\hbox to0pt{\hss$0$\hss}\vss}}
\put(5,10){\line(1,2){55}}
\put(61,122){\vbox to0pt{\vss\hbox to0pt{$\psi_c=\mathrm{const}_3$\hss}}}
\put(90,20){\line(-1,2){50}}
\put(39,122){\vbox to0pt{\vss\hbox to0pt{\hss$\psi_b=\mathrm{const}_2$}}}
\put(50,100){\circle*{3}}
\put(44,100){\vbox to0pt{\vss\hbox to0pt{\hss$2$}\vss}}
\put(60,80){\circle*{3}}
\put(60,73){\vbox to0pt{\hbox to0pt{\hss$3$}\vss}}
\put(55,80){\line(1,0){85}}
\put(142,80){\vbox to0pt{\vss\hbox to0pt{$\psi_a=\mathrm{const}_4$\hss}}}
\put(120,80){\circle*{3}}
\put(120,75){\vbox to0pt{\hbox to0pt{$4$\hss}\vss}}
\put(90,20){\line(1,2){40}}
\put(131,102){\vbox to0pt{\vss\hbox to0pt{$\psi_c=\mathrm{const}_6$\hss}}}
\put(110,60){\circle*{3}}
\put(105,60){\vbox to0pt{\vss\hbox to0pt{\hss$5$}\vss}}
\put(135,10){\line(-1,2){40}}
\put(94,92){\vbox to0pt{\vss\hbox to0pt{\hss$\psi_b=\mathrm{const}_5$\hss}}}
\put(130,20){\circle*3}
\put(130,26){\vbox to0pt{\vss\hbox to0pt{$6$\hss}}}
\end{picture}}
\end{figure}
It is easy to see that, for any other possible arrangement,
the point $\widehat\psi(T)$ lies between either
$\widehat\psi(T_1),\widehat\psi(T_6)$, or
$\widehat\psi(T_2),\widehat\psi(T_3)$, or
$\widehat\psi(T_4),\widehat\psi(T_5)$.\end{proof}

In the particular case $M'=M$, we come to the following
\begin{coro}
Let $M$ be a closed triangulated surface such that the holonomy
group of the canonical connection is trivial. Then the only
solutions of the black (or white) triangle equation on $M$ are
covariant constants, which form a two-dimensional space.
\end{coro}

%%%%%%%%%%%%%%%%%%%%%%%%%%%%%%%%%%%%%%%%%%%%%%%%%%%%%%%%%%%%%%%%%%%%%%
%%%%%%%%%%%%%%%%%%%%%%%%%%%%%%%%%%%%%%%%%%%%%%%%%%%%%%%%%%%%%%%%%%%%%%

\section{Triangle equations in the Euclidean geometry}
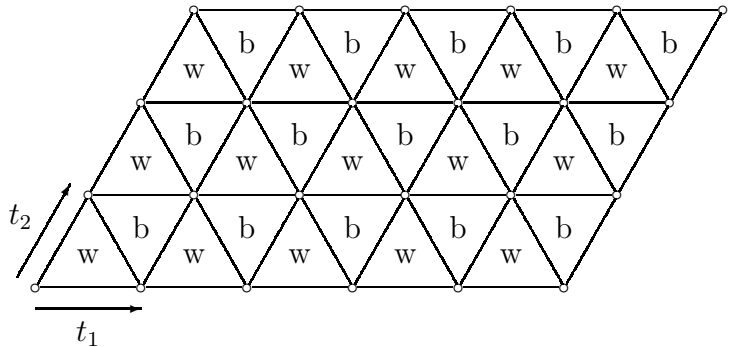
\begin{figure}[ht]\caption{Black and white coloring of
the Euclidean plane}\label{plane}\vskip5pt
\centerline{\begin{picture}(290,150)
\multiput(20,30)(20,35){3}{\begin{picture}(220,35)
\multiput(0,0)(40,0){5}{\begin{picture}(40,35)
\put(0,0){\circle{3}}
\put(2,0){\line(1,0){36}}
\bezier200(0.8,1.4)(10,17.5)(19.2,33.6)
\bezier200(39.2,1.4)(30,17.5)(20.8,33.6)
\put(20,12){\vbox to0pt{\vss\hbox to0pt{\hss w\hss}\vss}}
\put(40,23){\vbox to0pt{\vss\hbox to0pt{\hss b\hss}\vss}}
\end{picture}}
\put(200,0){\circle{3}}
\bezier200(200.8,1.4)(210,17.5)(219.2,33.6)
\end{picture}}
\multiput(80,135)(40,0){6}{\circle{3}}
\multiput(82,135)(40,0){5}{\line(1,0){36}}
\put(20,22){\vector(1,0){40}}
\put(40,17){\vbox to0pt{\hbox to0pt{\hss$t_1$\hss}\vss}}
\bezier200(13,34)(23,51.5)(33,69)
\put(33,69){\vector(1,2){0}}
\put(19,54){\vbox to0pt{\vss\hbox to0pt{\hss$t_2$}}}
\end{picture}}
\end{figure}

From now on, we will consider the case when our surface $M$ is the
Euclidean plane $\real^2$ triangulated so that all  triangles are
equilateral. Its vertices form a two-dimensional lattice
$\ell\subset\real^2$, $\ell\cong\integer^2$. We denote by $t_1$
and $t_2$ the basis shifts of the lattice.
Each triangle of this triangulation
is either of the form $\langle P,t_1\cdot P,t_2\cdot
P\rangle$ (in which case we call it \emph{white} and denote by $T_P^{\mathrm
w}$), or of the form $\langle P,t_1^{-1}\cdot P,t_2^{-1}\cdot
P\rangle$ (in which case we call it \emph{black} and denote by $T_P^{\mathrm
b}$), see Fig.~\ref{plane}.

Thus, in this case, there is a natural one-to-one correspondence
between vertices of the triangulation and black (white) triangles:
$P\leftrightarrow T_P^{\mathrm b}$ (respectively,
$P\leftrightarrow T_P^{\mathrm w}$). Due to this correspondence,
we can think of operators $Q_{\mathrm b},Q_{\mathrm w}$ of the
form~(\ref{1order}) associated with the set of black or white
triangles, respectively, as ones mapping the space of functions on
the lattice $\integer^2$ to itself. In other words, $Q_{\mathrm
b},Q_{\mathrm w}$ have the form
\begin{equation}
\begin{aligned}
Q_{\mathrm w}&=x_n+y_nt_1+z_nt_2,\\
Q_{\mathrm b}&=u_n+v_nt_1^{-1}+w_nt_2^{-1},
\end{aligned}
\end{equation}
where $n=(n_1,n_2)\in\integer^2$. Notice, that
operators of ``black'' type $Q_{\mathrm b}$ are (formally)
conjugated to operators of ``white'' type $Q_{\mathrm w}$,
as we have $t_j^+=t_j^{-1}$.

Consider a real   self-adjoint second order (Schr\"odinger)
operator
\begin{equation}
(L\psi)_P=\sum\limits_{P'}b_{P,P'}\psi_{P'},\quad
b_{P,P'}=b_{P',P},
\end{equation}
where $b_{P,P'}>0$ holds if either we have $P=P'$ or $\langle P,P'\rangle$
is an edge, and we have $b_{P,P'}=0$ otherwise. In other words,
$L$ is an operator of the form
\begin{equation}
L=a_n+b_nt_1+c_nt_2+d_nt_1^{-1}t_2+e_nt_1^{-1}+f_nt_2^{-1}+g_n
t_1t_2^{-1}.
\end{equation}
It admits a unique factorization of the form
\begin{equation}
L=Q_{\mathrm b}^+Q_{\mathrm b}+U_P
\end{equation}
and of the form
\begin{equation}
L=Q_{\mathrm w}^+Q_{\mathrm w}+V_P,
\end{equation}
where all coefficients of $Q_{\mathrm b}$ and $Q_{\mathrm w}$
are positive.

Interesting classes of such operators and their spectral theory in
the space ${\cal L}_2(\integer^2)$ are studied in \cite{nd,n1}. In
particular, the attention is paid to zero modes of the operator
$Q_{\mathrm b}^+Q_{\mathrm b}$ (respectively, $Q_{\mathrm
w}^+Q_{\mathrm w}$), which are functions $\psi\in{\cal
L}_2(\integer^2)$ such that $Q_{\mathrm b}\psi=0$ (respectively,
$Q_{\mathrm w}\psi=0$). Explicit solutions are found in the case
of coefficients of the form $\exp(l_j(n))$, where
$l_j(n)=l_{j1}n_1+ l_{j2}n_2$, $j=1,2$, are linear forms:
\begin{equation}
Q_{\mathrm w}=1+ce^{l_1(n)}t_1+de^{l_2(n)}t_2=Q(c,d).
\end{equation}
Especially interesting is the case when $l_{ij}+l_{ji}=h$
is independent of $i,j$. In this case, the operators
$Q(c,d)$ satisfy the relation
\begin{equation}
Q(c,d)^+Q(c,d)-1=q^2(Q(c/q^2,d/q^2)Q(c/q^2,d/q^2)^+-1),
\end{equation}
where $q=e^{l_{11}}=e^{l_{22}}=e^{(l_{12}+l_{21})/2}$. In some
cases, these relations allow to establish a wonderful property of
the spectrum of operators $L=Q^+Q$ and $L'=QQ^+$ in the Hilbert
space ${\cal L}_2(\integer^2)$ similar to that of the famous
``Landau operator'' in the magnetic field, though there is nothing
 like a physical magnetic field in our situation. This phenomena are
coming only from the nature of the discrete systems.

A pair of operators $(Q_{\mathrm b},Q_{\mathrm w})$ defines a
discrete connection on the plane $\real^2$ as described in
Section~1. As shown in \cite{nd}, this connection has trivial
local holonomy in the special case $l_{ij}+l_{ji}=0$.

It is also noticed in~\cite{nd} that
operators $Q_{\mathrm w}=1+y_nt_1+z_nt_2$
and $Q_{\mathrm b}=1+v_nt_1^{-1}+w_nt_n^{-1}$ define
a connection of zero curvature if and only if there exists
an everywhere non-zero function $f$ such that
\begin{equation}
((Q_{\mathrm w}-1)(Q_{\mathrm b}-1)-1)=
f\cdot((Q_{\mathrm b}-1)(Q_{\mathrm w}-1)-1).
\end{equation}

\section{Discrete analog of the operators $\partial$
and~$\overline\partial$}

Recall that we consider an equilateral triangular lattice
$\ell\cong\integer^2$ in the Euclidean plane $\real^2$ and the
corresponding b/w triangulation: the triangles $T_n^{\mathrm
w}=\langle(n_1,n_2),(n_1+1,n_2), (n_1,n_2+1)\rangle$ are white,
and the triangles $T_n^{\mathrm
b}=\langle(n_1,n_2),(n_1-1,n_2),(n_1,n_2-1)\rangle$ are black,
where $n=(n_1,n_2)\in\integer^2$.

We will consider the following operators on the
space of functions on $\ell$:
\begin{equation}
Q=1+t_1+t_2,\qquad Q^+=1+t_1^{-1}+t_2^{-1},
\end{equation}
which are formally conjugate to each other, and show that they
have many properties similar in a sense to those of the operators
$\partial=\partial/\partial z$ and
$\overline\partial=\partial/\partial\overline z$ on the complex
plane.

We denote by $H$ the kernel of the operator $Q^+$. It will
play a role of the space of \emph{holomorphic functions}. It
consists of functions $\psi$ on $\ell$ that have zero sum over all the
vertices of any black triangle: $Q^+\psi=0$.
We have not found any natural algebra structure on $H$,
but we have constructed analogs of polynomials, the Taylor expansion,
and the Cauchy formula. A discrete version of the maximum
principle, which was considered in Section~\ref{mp} in a
more general situation, also takes place.

Now we introduce the space ${\cal P}_k$
of ``polynomials of degree $\leqslant k$'',
which are solutions of the following system of equations:
\begin{equation}
Q^+\psi=0,\qquad Q^{k+1}\psi=0.
\end{equation}

\begin{prop}\label{dim2k}
The dimension of the subspace ${\cal P}_{k}$ equals $2k+2$.
\end{prop}

\begin{proof}
By definition, ${\cal P}_0$ is the space of functions $\psi$
satisfying both equations
\begin{equation}\label{af=0,a+f=0}
Q\psi=0,\qquad Q^+\psi=0.
\end{equation}
It means that $\psi$ is a covariant constant with respect
to the canonical connection of our triangulation. As we have seen
in Section~\ref{disconn}, in the case of trivial holonomy,
the space of covariant constants is two-dimensional. Thus, we
have
$$\dim{\cal P}_0=2.$$

Notice that, since the operators $Q$ and $Q^+$ commute, the
space $H$ of ``holomorphic'' functions is invariant under $Q$:
$Q(H)\subset H$. We claim that we also have~$Q(H)\supset H$,
{\it i.e.},~for any function $\varphi\in H$,
the following system of equations is consistent:
\begin{equation}\label{af=g,a+f=0}
Q\psi=\varphi,\quad Q^+\psi=0.
\end{equation}
Indeed, the system defines an \emph{affine} discrete connection
on the plane similar to the canonical connection
from Section~\ref{disconn}. One can easily check that
the local holonomy of this connection is trivial
if and only if we have $Q^+\varphi=0$, which
holds by assumption. The rest of the argument
is the same as in the case of linear discrete connection.

Thus, the restriction of $Q$ to the space $H$ is
a linear operator whose image is the whole space $H$
and the kernel is two-dimensional. It follows
immediately that
$$\dim({\cal P}_{k+1}/{\cal P}_k)=2.$$\end{proof}

\begin{rem}
One can show that, for any $\psi\in{\cal P}_k\backslash{\cal P}_{k-1}$,
there exists a degree $k-1$ polynomial $u_{k-1}(z,\overline z)$
with complex coefficients such that the following holds
\begin{equation}
\psi_n=\qopname\relax o{Re}\left(\alpha e^{\frac{2\pi i}3(n_1-n_2)}
(z^k+u_{k-1}(z,\overline z))\right),\end{equation}
where $n=(n_1,n_2)$, $z=n_1+e^{\frac{2\pi i}3}n_2$, $\alpha\in\mathbb C$.
\end{rem}

Now we introduce an analog of the Taylor expansion
for functions from $H$. We will need to consider
``big black triangles'' $T_n^{(k)}=
\langle(n_1,n_2),(n_1-2k-1,n_2),(n_1,n_2-2k-1)\rangle$,
where $n=(n_1,n_2)\in\ell$.
A big black triangle is homothetic to an ordinary
black triangle and has $2k+2$ vertices of the lattice in each side.

We need the following preparatory lemma.

\begin{lemm}\label{interpolation}
For any ``holomorphic function'' $\psi\in H$, any
$n\in\ell$ and $k\geqslant0$, there exists
a unique ``degree $\leqslant k$
polynomial'' $p_k\in{\cal P}_k$ such that
$p_k$ coincides with $\psi$ in $T_n^{(k)}$.
\end{lemm}

\begin{proof}
We prove the existence by induction.
For $k=0$, the assertion is true, since
there always exists a covariant constant $p_0$ coinciding
with $\psi$ in just one black triangle.

Assume that the assertion is proved for $k=l-1$. Then we
can find $p_{l-1}\in{\cal P}_{l-1}$ such that $p_{l-1}=Q\psi$ wherever
in $T^{(l-1)}_{(n_1-1,n_2-1)}$. As we have already
seen, we can find a function $\varphi\in H$
such that $Q\varphi=p_{l-1}$, and a covariant constant
$p_0\in{\cal P}_0$ such that $p_0=\psi-\varphi$ in
$T_n^{\mathrm b}=T_n^{(0)}$. We set $p_l=\varphi+p_0$.
By construction, we will have: $p_l\in{\cal P}_l$,
$p_l=\psi$ in $T_n^{(0)}$ and $Qp_l=Q\psi$ in
$T_{(n_1-1,n_2-1)}^{(l-1)}$.

Thus, the difference $p_l-\psi$ satisfies the following
$$Q^+(p_l-\psi)=0\mbox{ everywhere},\quad
Q(p_l-\psi)=0\mbox{ in }T^{(l-1)}_{(n_1-1,n_2-1)},$$
which means that the sum of the values of the function $p_l-\psi$
over the vertices of any triangle, black or white, contained in
$T^{(l)}_n$ is zero. Since $p_l=\psi$ in $T_n^{(0)}$,
this implies that the function $p_l-\psi$
is identically zero in $T^{(l)}_n$.

The uniqueness of the ``polynomial'' $p_k$ follows from the dimension
argument: the dimension of ${\cal P}_k$ is $2k+2$ by Proposition~\ref{dim2k},
and the $2k+2$ values of $\psi$ at vertices lying
in one side of $T_n^{(k)}$, say, points $(n_1-j,n_2)$, where
$j=0,1,\dots,2k+1$, can be arbitrary as implied by the
following lemma.\end{proof}

\begin{lemm}\label{trefoil}
Let $Y_{n_1n_2}$ be the following subset of the lattice $\ell$:
\begin{equation}
Y_n=\{(n_1,n_2)\}\cup\{(n_1-j,n_2),\ (n_1,n_2+j),\
(n_1+j,n_2-j)\}_{j\geqslant 1},
\end{equation}
where $n=(n_1,n_2)$. Let $V_n$ be the linear space of
all real functions on $Y_n$.
Then, for any $n\in\ell$ the restriction map
$H\rightarrow V_n$ is an isomorphism.
\end{lemm}

In other words, in order to get a solution to the equation
$Q^+\psi=0$ one can set values of $\psi$ in all points from $Y_n$
arbitrarily, and then the equation will prescribe the values of
$\psi$ in all the other points.

\begin{proof}
The three rays with the origin $n$ at which all the points from
$Y_n$ lie, cut the plane into three parts, which are cyclically
interchanged under rotation by $2\pi/3$ around $n$. Consider one
of these parts $U=\{(n_1-j_1,n_2+j_2)\}_{j_1,j_2>0}$. All black
triangles in $U$ have the form $T^{\mathrm
b}_{(n_1-j_1+1,n_2+j_2)}$, where $j_1,j_2>0$. The corresponding
equations are:
\begin{equation}
\psi_{n_1-j_1,n_2+j_2}=
-\psi_{n_1-(j_1-1),n_2+j_2}-\psi_{n_1-(j_1-1),n_2+(j_2-1)}.
\end{equation}
They allow to express recursively the value of $\psi$ at
any point from $\ell\cap U$ via the values of $\psi$ at
$\ell\cap\partial U$.

A similar situation takes place in the other two parts of the plane
that are obtained from $U$ by rotation by $2\pi/3$ around
$n$. The recursion process is illustrated in Fig.~\ref{recur},
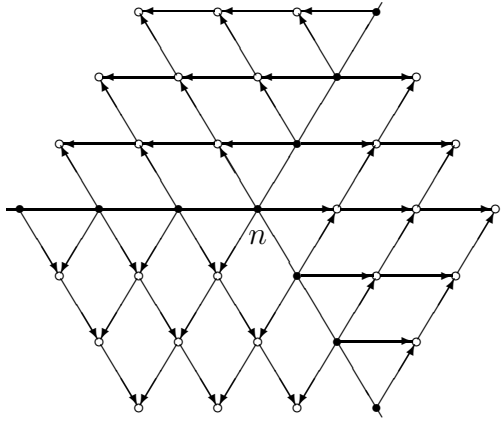
\begin{figure}[ht]\caption{Recursive construction of $\psi$}\label{recur}
\vskip10pt
\centerline{\begin{picture}(190,160)
\multiput(5,80)(30,0){4}{\circle*3}
\multiput(110,105)(15,25){3}{\circle*3}
\multiput(110,55)(15,-25){3}{\circle*3}
\put(0,80){\line(1,0){95}}
\put(95,80){\line(3,5){47}}
\put(95,80){\line(3,-5){47}}
\multiput(95,80)(15,25){3}{\begin{picture}(0,0)
\multiput(0,0)(-30,0){3}{\begin{picture}(0,0)
\put(-0.6,1){\line(-3,5){13.8}}
\put(-15,25){\circle{3}}
\put(-14.4,24){\vector(-1,2){0}}
\put(13.5,25){\vector(-1,0){27}}
\end{picture}}
\end{picture}}
\multiput(95,80)(15,-25){3}{\begin{picture}(0,0)
\multiput(0,0)(-30,0){3}{\begin{picture}(0,0)
\put(-0.6,-1){\line(-3,-5){13.8}}
\put(-14.4,-24){\vector(-1,-2){0}}
\put(-29.4,-1){\line(3,-5){13.8}}
\put(-15.6,-24){\vector(1,-2){0}}
\put(-15,-25){\circle{3}}
\end{picture}}
\end{picture}}
\multiput(95,80)(15,25){3}{\begin{picture}(0,0)
\multiput(0,0)(15,-25){3}{\begin{picture}(0,0)
\put(1.5,0){\vector(1,0){27}}
\put(15.6,-24){\line(3,5){13.8}}
\put(30,0){\circle{3}}
\put(29.4,-1){\vector(1,2)0}
\end{picture}}
\end{picture}}
\put(95,72){\vbox to0pt{\hbox to0pt{\hss$n$\hss}\vss}}
\end{picture}}
\end{figure}
where the vertices from $Y_n$ are marked by bold circles.
\end{proof}

\begin{figure}[ht]\caption{``Polynomials''
$p_{k,1}$
in $T^{(k)}_n$; $k=1,2$}\label{pk12}\vskip10pt
\centerline{\begin{picture}(200,90)
\put(0,30){\begin{picture}(50,40)
\put(25,10){\circle3}
\put(10,35){\circle3}
\put(40,35){\circle3}
\put(12,35){\line(1,0){26}}
\put(25.6,11){\line(3,5){13.8}}
\put(24.4,11){\line(-3,5){13.8}}
\put(30,7){\vbox to0pt{\vss\hbox to0pt{$1$\hss}\vss}}
\put(45,32){\vbox to0pt{\vss\hbox to0pt{$-1$\hss}\vss}}
\put(5,35){\vbox to0pt{\vss\hbox to0pt{\hss$0$}\vss}}
\end{picture}}
\put(90,0){\begin{picture}(110,90)
\multiput(10,85)(30,0){4}{\circle3}
\multiput(10,78)(30,0){3}{\vbox to0pt{\hbox to0pt{\hss$0$\hss}\vss}}
\multiput(12,85)(30,0){3}{\line(1,0){26}}
\multiput(25,60)(30,0){3}{\circle3}
\multiput(25,53)(30,0){2}{\vbox to0pt{\hbox to0pt{\hss$0$\hss}\vss}}
\multiput(25.6,61)(30,0){3}{\line(3,5){13.8}}
\multiput(24.4,61)(30,0){3}{\line(-3,5){13.8}}
\multiput(27,60)(30,0){2}{\line(1,0){26}}
\multiput(40,35)(30,0){2}{\circle3}
\put(40,28){\vbox to0pt{\hbox to0pt{\hss$0$\hss}\vss}}
\multiput(40.6,36)(30,0){2}{\line(3,5){13.8}}
\multiput(39.4,36)(30,0){2}{\line(-3,5){13.8}}
\put(42,35){\line(1,0){26}}
\put(55,10){\circle3}
\put(55.6,11){\line(3,5){13.8}}
\put(54.4,11){\line(-3,5){13.8}}
\multiput(55,10)(30,50){2}{\vbox to0pt{\hbox to0pt{$-1$\hss}\vss}}
\multiput(70,35)(30,50){2}{\vbox to0pt{\hbox to0pt{$\phantom-1$\hss}\vss}}
\end{picture}}
\end{picture}}
\end{figure}
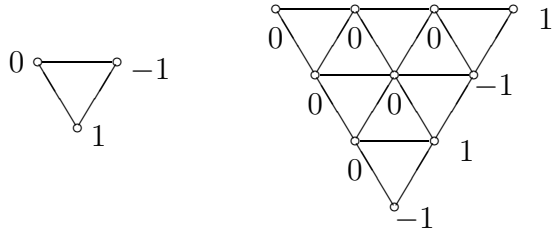

There is a canonical way to associate three ``polynomials of
degree $k$'' $p_{k,1},p_{k,2},p_{k,3}$
with any big black triangle $T_n^{(k)}$ by setting them
to be identically zero everywhere in $T_n^{(k)}$ except
at vertices lying in one of the sides where
they are equal to $\pm1$, namely:
\begin{equation}
\begin{aligned}
{}&p_{k,1}(n_1-2k-1+j,n_2)&&=(-1)^{j+k};\\
{}&p_{k,2}(n_1,n_2-j)&&=(-1)^{j+k};\\
{}&p_{k,3}(n_1-j,n_2-2k-1+j)&&=(-1)^{j+k},
\end{aligned}
\end{equation}
$j=0,1,\dots,2k+1$.
In Fig.~\ref{pk12} the values of $p_{k,1}$ in $T_n^{(k)}$
are shown for $k=1,2$.
The pictures for $p_{k,2}$ and $p_{k,3}$ are
obtained by rotation by $2\pi/3$.
Obviously, we have
\begin{equation}
Qp_{k,i}=p_{k-1,i},
\end{equation}
where $p_{k,i}$ is associated with $T_{(n_1,n_2)}^{(k)}$,
and $p_{k-1,i}$ with $T_{(n_1-1,n_2-1)}^{(k-1)}$.

The polynomials $p_{k,j}$ are linearly dependent:
\begin{equation}
p_{k,1}+p_{k,2}+p_{k,3}\in{\cal P}_{k-1},
\end{equation}
where we assume that the polynomials are associated with
the same big triangle $T_n^{(k)}$. Thus, in order
to get a basis in the space of polynomials we
have to select a big triangle $T_n^{(k)}$ for each
$k$ and a pair of polynomials $p_{k,i},p_{k,j}$, $i,j\in\{1,2,3\}$.
We do it as follows.

\begin{figure}[ht]\caption{Two-side extensions}\label{ext}
$$\begin{array}{ccc}
\begin{picture}(80,65)
\put(40,10){\line(3,5){30}}
\put(43,5){\line(-3,5){36}}
\put(43,5){\line(3,5){36}}
\put(10,60){\line(1,0){60}}
\put(7,65){\line(1,0){72}}
\end{picture}&\hskip20pt
\begin{picture}(80,65)
\put(37,5){\line(3,5){36}}
\put(40,10){\line(-3,5){30}}
\put(37,5){\line(-3,5){36}}
\put(10,60){\line(1,0){60}}
\put(1,65){\line(1,0){72}}
\end{picture}\hskip20pt&
\begin{picture}(80,65)
\put(40,10){\line(3,5){30}}
\put(40,0){\line(3,5){36}}
\put(40,10){\line(-3,5){30}}
\put(40,0){\line(-3,5){36}}
\put(4,60){\line(1,0){72}}
\end{picture}\\
\mbox{type (12)}&\mbox{type (23)}&\mbox{type (31)}
\end{array}$$
\end{figure}
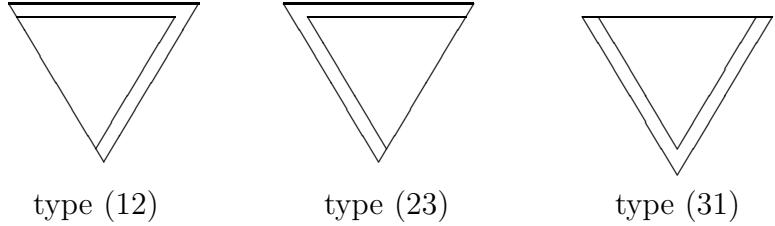

For a big triangle $T_{(n_1,n_2)}^{(k)}$, we define
its \emph{two-side extension} of type $(12)$, $(23)$,
or $(31)$ to be the big triangle $T_{(n_1+1,n_2+1)}^{(k+1)}$,
$T_{(n_1+1,n_2)}^{(k+1)}$, or $T_{(n_1,n_2+1)}^{(k+1)}$,
respectively (see Fig.~\ref{ext}).

Let us fix a sequence of big triangles
$T(0),T(1),\dots$ such that
\begin{enumerate}
\item $T(0)$ is an ordinary black triangle;
\item $T(k+1)$ is a two-side extension of $T(k)$ (thus,
$T(k)=T_n^{(k)}$ for some $n\in\ell$);
\item the union $\bigcup_kT(k)$ is the whole plane $\real^2$.
\end{enumerate}
Such a sequence will be called \emph{admissible}.

With an admissible sequence of triangles we
associate a basis $\psi^1_j,\psi^2_j$,
$j=0,1,\dots$, in the space $\bigoplus_k{\cal P}_k$
of polynomials in the following way.
If $T(k)$ is the two-side extension of $T(k-1)$ of type
$(ij)$, we set $\psi^1_k$ and $\psi^2_k$ to
be the polynomials $p_{k,i}$ and $p_{k,j}$ associated
with the triangle $T(k)$. We also denote by $T^{\mathrm b}(k)$
the (ordinary) black triangle $T_{(n_1-k,n_2-k)}^{\mathrm b}$,
where $T(k)=T_{(n_1,n_2)}^{(k)}$.

By construction, we have $\psi^i_j=0$ everywhere in $T(k)$
if $k<j$. Thus, any series of the form
\begin{equation}\label{taylor}
\sum\limits_{k=0}^\infty(\alpha_k^1\psi_k^1+\alpha_k^2\psi_k^2)
\end{equation}
converges everywhere in $\ell$, since, by the definition of
admissible sequence, for any $n\in\ell$,
we have $n\in T(k)$ for any sufficiently large $k$.

\begin{theo}
For any admissible sequence of triangles $T(0),T(1),\dots$
and any ``holomorphic'' function $\psi\in H$,
there exists a unique series of coefficients $\alpha^1_0,\alpha^2_0,
\alpha^1_1,\alpha^2_1,\alpha^1_2,\alpha^2_2,\dots$
such that the series~(\ref{taylor}) converges
to the function $\psi$. Moreover, the coefficients
$\alpha_k^1,\alpha_k^2$ can be found from the knowledge
of the value of the \emph{``$k$th derivative''} $Q^k\psi$
at the vertices of the triangle $T^{\mathrm b}(k)$.
\end{theo}

\begin{proof}
The first assertion of the theorem follows from the fact
that we can approximate $\psi$ in the triangle $T(k)$
by a ``polynomial'' of degree $k$ and from $\lim\limits_{k\rightarrow
\infty}T(k)=\real^2$.

It is straightforward to check that, if $\psi$ is identically
zero in $T(k)$, then $Q^k\psi$ is identically zero in $T^{\mathrm b}(k)$.
Thus, $Q^j\psi_k^i$ is not identically zero in $T^{\mathrm b}(k)$
if and only if $j=k$. This implies the latter assertion of the
theorem.\end{proof}

Now we construct an analog of the Cauchy formula
for recovering a solution $\psi$ of the equation $Q^+\psi=0$
in a domain from the knowledge of $\psi$ at the boundary
of the domain.

By a domain we will always mean a closed
domain $D$ in the plane such that $D$ is a simplicial
subcomplex of our regular triangulation.
For a black or white triangle $T$, we will write
$T\in\partial_+D$ if we have $T\not\subset D$ and $T\cap\partial D
\ne\varnothing$.

Consider the following function (the famous Pascal triangle, see
Fig.~\ref{pasc}).
\begin{equation}\label{green}
G_{(n_1,n_2)}=\left\{\begin{array}{ll}
(-1)^{n_1+n_2}\begin{pmatrix} n_1\\n_1+n_2\end{pmatrix} &\mbox{if
}n_1,n_2\geqslant0,\\ 0&\mbox{otherwise}.
\end{array}\right.
\end{equation}
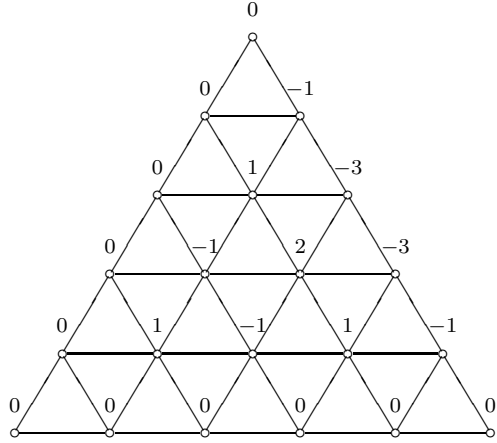
\begin{figure}\caption{The function G}\label{pasc}
\centerline{\begin{picture}(180,170)
\multiput(0,0)(36,0){5}{\begin{picture}(0,0)
\put(0,0){\circle{3}}\put(2,0){\line(1,0){32}}\put(0.6,1){\line(3,5){16.8}}
\put(35.4,1){\line(-3,5){16.8}}
\end{picture}}
\multiput(18,30)(36,0){4}{\begin{picture}(0,0)
\put(0,0){\circle{3}}\put(2,0){\line(1,0){32}}\put(0.6,1){\line(3,5){16.8}}
\put(35.4,1){\line(-3,5){16.8}}
\end{picture}}
\multiput(36,60)(36,0){3}{\begin{picture}(0,0)
\put(0,0){\circle{3}}\put(2,0){\line(1,0){32}}\put(0.6,1){\line(3,5){16.8}}
\put(35.4,1){\line(-3,5){16.8}}
\end{picture}}
\multiput(54,90)(36,0){2}{\begin{picture}(0,0)
\put(0,0){\circle{3}}\put(2,0){\line(1,0){32}}\put(0.6,1){\line(3,5){16.8}}
\put(35.4,1){\line(-3,5){16.8}}
\end{picture}}
\put(72,120){\begin{picture}(0,0)
\put(0,0){\circle{3}}\put(2,0){\line(1,0){32}}\put(0.6,1){\line(3,5){16.8}}
\put(35.4,1){\line(-3,5){16.8}}
\end{picture}}
\multiput(180,0)(-18,30)6{\circle{3}}
\multiput(0,8)(36,0){6}{\hbox to 0pt{\hss$\scriptstyle0$\hss}}
\multiput(18,38)(18,30){5}{\hbox to 0pt{\hss$\scriptstyle0$\hss}}
\put(54,38){\hbox to 0pt{\hss$\scriptstyle1$\hss}}
\put(90,38){\hbox to 0pt{\hss$\scriptstyle-1$\hss}}
\put(126,38){\hbox to 0pt{\hss$\scriptstyle1$\hss}}
\put(162,38){\hbox to 0pt{\hss$\scriptstyle-1$\hss}}
\put(72,68){\hbox to 0pt{\hss$\scriptstyle-1$\hss}}
\put(108,68){\hbox to 0pt{\hss$\scriptstyle2$\hss}}
\put(144,68){\hbox to 0pt{\hss$\scriptstyle-3$\hss}}
\put(90,98){\hbox to 0pt{\hss$\scriptstyle1$\hss}}
\put(126,98){\hbox to 0pt{\hss$\scriptstyle-3$\hss}}
\put(108,128){\hbox to 0pt{\hss$\scriptstyle-1$\hss}}
\end{picture}}
\end{figure}

\begin{lemm}
The function $G$ is a fundamental solution to the black triangle
equation, i.e.,~we have
\begin{equation}\label{ag=delta}
Q^+G=\delta,
\end{equation}
where $\delta$ is the delta-function
$$\delta_n=\left\{\begin{aligned}
1&\quad\mbox{if }n=(0,0),\\
0&\quad\mbox{otherwise}.
\end{aligned}\right.$$
\end{lemm}

\begin{proof}
The formula~(\ref{green}) can be rewritten in the form
\begin{equation}\label{green1}
G_n=\sum\limits_{k=0}^\infty
\left((-t_1^{-1}-t_2^{-1})^k\delta\right)_n.
\end{equation}
Applying $Q^+=(1+t_1^{-1}+t_2^{-1})$ to both sides of~(\ref{green1})
we obtain~(\ref{ag=delta}).\end{proof}

\begin{prop}\label{recover}
Let $\psi$ be a ``holomorphic function'' in a finite domain $D$,
i.e.,~a solution of the equation $Q^+\psi=0$ in $D$.
Then, for any $n\in D\cap\ell$, the following holds
\begin{equation}\label{cauchy}
\psi_n=\sum\limits_{T_m^{\mathrm b}\in\partial_+D}(Q^+\psi)_mG_{n-m},
\end{equation}
where we set $\psi_m=0$ for all $m\notin D$.
\end{prop}

Note that the right hand side of~(\ref{cauchy}) involves only
the values of $\psi$ at the boundary $\partial D$.

\begin{proof}
Notice that, if $T_m^{\mathrm b}\notin\partial_+D$, then we have
$(Q^+\psi)_m=0$. Indeed, if $T^{\mathrm b}_m\subset D$, then the equality
holds by assumption, and if $T^{\mathrm b}_m\cap D=\varnothing$,
it holds, since $\psi$ is assumed to be identically zero outside $D$.
So, the right hand side of~(\ref{cauchy}) can be
written as
\begin{equation}
\sum\limits_{m\in\ell}(Q^+\psi)_mG_{n-m},
\end{equation}
Denote this sum by $\widetilde\psi_n$.

We may assume without loss of generality that the domain $D$
is contained in the sector $n_1,n_2>1$. Then, for any $n=(n_1,n_2)$
such that $T_n^{\mathrm b}\in\partial_+D$, we have
$n_1,n_2>0$. This implies
\begin{equation}\label{f=0}
\widetilde\psi_n=\psi_n=0\quad\mbox{if }n_1\leqslant0\mbox{ or }
n_2\leqslant0.
\end{equation}

From~(\ref{ag=delta}) we have
$$\begin{aligned}
(Q^+(\psi-\widetilde \psi))_n&=
(Q^+\psi)_n-\sum\limits_{m\in\ell}(Q^+\psi)_m(Q^+G)_{n-m}\\
&=(Q^+\psi)_n-\sum\limits_{m\in\ell}(Q^+\psi)_m\delta_{n-m}\\
&=(Q^+\psi)_n-(Q^+\psi)_n=0.
\end{aligned}$$

Thus, the function $\psi-\widetilde\psi$ belongs to $H$ and,
according to~(\ref{f=0}), vanishes at $Y_0$. It follows
from Lemma~\ref{trefoil} that $\psi-\widetilde\psi=0$.
\end{proof}

In this proof, we have used two properties of the function $G$:
the first one is relation~(\ref{ag=delta}), and the second
one is that $G=0$ outside the sector $n_1,n_2\geqslant0$. However,
the latter is not needed as will follow from the
next lemma.

\begin{lemm}
For any function $\psi:\ell\rightarrow{\mathbb R}$
with finite support and any $\varphi\in H$ we have
\begin{equation}
\sum\limits_{m\in\ell}(Q^+\psi)_m\varphi_{n-m}=0
\end{equation}
for all $n\in\ell$.
\end{lemm}

\begin{proof}
Denote by $\tau$ the following operator:
$$(\tau\psi)_n=\psi_{-n}.$$
We will have
$$\begin{aligned}
\sum\limits_{m\in\ell}(Q^+\psi)_m\varphi_{n-m}&=
\sum\limits_{m\in\ell}(Q^+\psi)_m(\tau\varphi)_{m-n}\\
&=\sum\limits_{m\in\ell}\psi_m(Q\tau\varphi)_{m-n}\\
&=\sum\limits_{m\in\ell}\psi_m(\tau Q^+\varphi)_{m-n}=0.
\end{aligned}$$
We used here the relation $\tau Q\tau=Q^+$ and the fact
that the operators $Q$ and $Q^+$
are formally conjugate to each other.
\end{proof}

\begin{coro}
The assertion of Proposition~\ref{recover} takes place for
an arbitrary function $G$ satisfying~(\ref{ag=delta}).
\end{coro}

\section*{Appendix~1}
\begin{proof}[Proof of Theorem~\ref{anyrho}]
Any representation $\rho:\pi_1(M)\rightarrow GL(2,\real)$
can be obtained by specifying a flat linear connection
in the trivial two-dimensional vector bundle over
the 1-skeleton of $\cal T$. This means that, to each
oriented edge $\langle P,P'\rangle$, we associate a $2\times2$ matrix
$R_{P,P'}$ so that, for any triangle $\langle P,P',P''\rangle$
of $\cal T$, we have $R_{P,P''}=R_{P,P'}R_{P',P''}$ and
$R_{P,P'}=R_{P',P}^{-1}$. For a fixed vertex $P_0$,
a representation of $\pi_1(M,P_0)$ is obtained by
putting $\rho(\gamma)=R_{P_0P_1}\cdot\ldots\cdot R_{P_{m-1}P_m}R_{P_mP_0}$,
where the path $\gamma$ consists of the edges $\langle P_0,P_1\rangle,
\dots,\langle P_{m-1},P_m\rangle,\langle P_m,P_0\rangle$.
Clearly, any representation can be obtained in this way.
Moreover, the representation is not changed under
a ``gauge transformation'' $R_{P,P'}\mapsto C_PR_{P,P'}C_{P'}^{-1}$,
where $C_P$, $P\ne P_0$, are arbitrary $2\times2$
matrices, and $C_{P_0}=\mathrm{id}$.

Let us fix a vertex $P_0'$ connected with $P_0$ by an edge of $\cal T$.
By a gauge transformation, we can achieve $R_{P_0P_0'}=\begin{pmatrix}
0&1\\1&0\end{pmatrix}$. Let $D$ be a simply connected domain containing
$P_0,P_0'$. Consider a local covariantly constant
section of the vector bundle, {\it i.e.},
a vector function $v_P=(v_{P1},v_{P2})$
on vertices from $D$ such that $v_P\cdot R_{P,P'}=v_{P'}$ for
any edge $\langle P,P'\rangle\subset D$. Since we have
$v_{P_02}=v_{P_0'1}$, the section $v_P$ can be recovered
from the knowledge of $v_{P_01},v_{P_0'1}$. The space of
covariantly constant local sections is two-dimensional.

By forgetting the second coordinate, we obtain
a two-dimensional space of functions $v_{P1}$,
which can be turned into the space of covariant constants
of a flat discrete connection associated with $\cal T$.
This can be done by putting
\begin{equation}\label{coef}
b_{T,P_1}=c_{23},\quad b_{T,P_2}=c_{31}d_{23},\quad
b_{T,P_3}=c_{12}d_{21},
\end{equation}
where $T=\langle P_1,P_2,P_3\rangle$ is a triangle of $\cal T$
with a fixed enumeration of vertices,
$c_{ij}=(R_{P_i,P_j})_{21}$, $d_{ij}=\det(R_{P_iP_j})$.
Changing the enumeration will result in the multiplication
of all three coefficients~(\ref{coef}) by the same constant,
which has no effect on the corresponding discrete connection.
Applying a generic gauge transformation with $C_{P_0}=\mathrm{id}$
we can always make all the coefficients~(\ref{coef}) be non-zero.

The discrete connection $\{b_{T,P}\}$
obtained in this way has the following property.
If we set $\psi_{P_0},\psi_{P_0'}$ arbitrarily and extend $\psi_P$
by solving equation~(\ref{geneq}) along a thick path,
we will obtain the same function as
if we set $v_{P_0}=(\psi_{P_0},\psi_{P_0'})$,
apply the parallel transport along the same path by using
the connection $R$, and then forget
the second coordinate of the obtained vector function.
\end{proof}

\section*{Appendix~2}
Here we exhibit a natural generalization of the constructions of
Section~\ref{disconn}.

Let $X$ be a simplicial complex, $k\geqslant1$ an integer.
With any family $\cal K$ of $k$-dimensional simplices and
numeric non-zero coefficients $b_{\sigma^k,P}$ defined for
any simplex $\sigma^k\in\cal K$ and its vertex $P\in\sigma^k$
we associate the operator
\begin{equation}
(Q\psi)_{\sigma^k}=\sum\limits_{P\in\sigma^k}b_{\sigma^k,P}\psi_P,
\end{equation}
which takes functions of a vertex to functions on $\cal K$.

An analogue of discrete connections of Section~\ref{disconn} can
be defined if $\cal K$ is the set of all $k$-simplices of $X$
and $X$ satisfies certain restrictions, which we now describe.

By a $k$-\emph{thick path} in $X$ we shall mean a sequence
$\sigma_1,\sigma_2,\dots,\sigma_m$ of $k$-dimensional simplices
such that $\sigma_i\cap \sigma_{i+1}$ is a $(k-1)$-dimensional
simplex for any $i=1,\dots,m-1$. We call a thick path
$\sigma_1,\dots,\sigma_m$ \emph{elementary} if all simplices
$\sigma_i$ belong to the star of a single vertex. If a thick
path $\gamma_1$ is obtained from a thick path $\gamma_2$
by an insertion or a removal an elementary path we say
that the passage $\gamma_1\mapsto\gamma_2$
is an \emph{elementary homotopy}.

Any thick path $\sigma_1,\dots,\sigma_m$
defines a homotopy class of ordinary paths from
the center of $\sigma_1$ to the center of $\sigma_m$
in the obvious way. Thus, it makes sense to speak
about homotopic thick paths.

We say that a simplicial complex $X$ is $k$-\emph{admissible}
if any path is $k$-``thickable'' and any homotopy
between $k$-thick paths can be realized as
a composition of elementary homotopies. For example,
the $k$-skeleton of a triangulated manifold is always $k$-admissible.

For any $k$-thick loop
$\gamma=(\sigma_1,\dots,\sigma_m,\sigma_{m+1}=\sigma_1)$
we define the \emph{holonomy operator} $R_\gamma$ in
the same way as we did in Section~\ref{disconn}.
$R_\gamma$ is a linear operator on the space of solutions
of the equation $Q\psi=0$ in the simplex $\sigma_1$.
We say that the connection $\{b_{\sigma,P}\}$ has
\emph{trivial local holonomy} (or \emph{zero curvature})
if $R_\gamma=\mathrm{id}$ for any elementary $k$-thick loop $\gamma$.

\begin{lemm}
If $X$ is a $k$-admissible complex, then for any
$k$-simplex $\sigma^k_0$ the global holonomy of
any discrete connection $\{b_{\sigma^k,P}\}$ with zero curvature
along $k$-thick loops based at $\sigma^k_0$
is a well defined homomorphism $\pi_1(X,\sigma^k_0)\rightarrow
\real^k$.
\end{lemm}

The proof is easy.

\medskip

Consider the canonical connection associated with the family of
all $k$-simplices of $X$, $b_{\sigma^k,P}\equiv1$.
It is easy to see that the canonical connection
necessarily has trivial local holonomy at any vertex
in the following two cases:
\begin{enumerate}
\item
$k=1$;
\item
$k\geqslant2$, $X$ is a triangulated $k$-manifold such that
any $(k-2)$-dimensional simplex $\sigma^{k-2}$
({\it i.e.}, the number of $k$-simplices containing $\sigma^{k-2}$)
is even.
\end{enumerate}
We will assume one of these in the sequel.
The operator $L=Q^+Q$ has the form
$$(L\psi)_P=\sum\limits_{P'}m_{P,P'}\psi_{P'}+n_P\psi_P,$$
where the sum is taken over all vertices $P'\ne P$ from the
star of $P$, $m_{P,P'}$ is the number of $k$-simplices
containing the edge $\langle PP'\rangle$,
$n_P$ is the number of $k$-simplices containing the vertex $P$.

By analogy with Propositions~\ref{p1}, \ref{p2} one can prove the following.

\begin{prop}
The holonomy group $\cal G$ of the canonical connection of a
triangulated $k$-manifold $X$ all whose $(k-2)$-simplices have even valence
is a subgroup of the permutation group $S_{k+1}$.

Let $q$ be the number of orbits of the action of ${\cal G}\subset S_{k+1}$
on the set $\{0,1,\dots,k\}$. Then the dimension of the space
of covariant constants of the canonical connections, i.e.,
solutions of the equation $Q\psi=0$ is $(q-1)$-dimensional.

The space of covariant constants coincides with the space
of zero modes of the operator $L$.
\end{prop}

In the particular case $k=1$, zero modes of the operator $L$ exist
if and only if the 1-skeleton of $X$ is a dichromatic graph.
Now we suppose $k\geqslant2$.

Consider the following three homomorphisms $\pi_1(X)\rightarrow\integer_2$:

\begin{enumerate}
\item
the parity $\rho_1$ of the global holonomy of the canonical connection;
\item
the orientation homomorphism $\rho_2$;
\item
the parity homomorphism $\rho_3$ of the number of $k$-simplices
in a $k$-thick path.
\end{enumerate}

By analogy with Lemmas~\ref{l2}, \ref{l3}, we have the following.

\begin{lemm}
A triangulated $k$-manifold all whose $(k-2)$-simplices
have even valence admits a b/w coloring of $k$-simplices
such that any two adjacent $k$-simplices are of different color
if and only if the representation $\rho_3$ is trivial.

We have $\rho_3=\rho_1\rho_2$.
\end{lemm}

If $X$ admits a b/w coloring of $k$-simplices, then the operator
$L$ can also be factorized as follows:
$$L=2Q_{\mathrm b}^+Q_{\mathrm b}=2Q_{\mathrm w}^+Q_{\mathrm w},$$
where $Q_{\mathrm b},Q_{\mathrm w}$ are operators associated
with the family of black and white simplices, respectively.
It is natural to ask whether the maximum principle holds
for these operators. This question has not yet been investigated.

\end{document}